\documentclass[twocolumn,aps,prb,showpacs,tightenlines,amsmath,amssymb]{revtex4}
\usepackage{graphicx}    
\usepackage{amssymb}  
\usepackage{dcolumn}
\usepackage{amsmath}     
\usepackage{bm}                                          
\usepackage{colordvi}                   

\begin{document}             
\title{Strongly modulated transmissions in gapped armchair graphene nanoribbons
  with sidearm or on-site gate voltage}
\author{H. Tong}
\affiliation{Hefei National Laboratory for Physical Sciences at Microscale and
  Department of Physics, University of Science and Technology of China, Hefei,
  Anhui, 230026, China} 
\author{M. W. Wu}
%\thanks{Author to  whom correspondence should be addressed}
\email{mwwu@ustc.edu.cn.}
\affiliation{Hefei National Laboratory for Physical Sciences at Microscale and
  Department of Physics, University of Science and Technology of China, Hefei,
  Anhui, 230026, China} 
\date{\today}

\begin{abstract}
We propose two schemes of field-effect transistor based on gapped armchair
graphene nanoribbons connected to metal leads, by introducing sidearms or
on-site gate voltages. We make use of the band
gap to reach excellent switch-off character. By introducing one sidearm or
on-site gate to the graphene nanoribbon, conduction peaks appear inside the
gap regime. By further applying two sidearms or on-site gates, these peaks
are broadened to conduction plateaus with a wide energy window, thanks to the
resonance from the dual structure. The position of the conduction windows inside
the gap can be fully controlled by the length of the sidearms or the on-site gate
voltages, which allows ``on'' and ``off'' operations for a specific energy window inside the
gap regime. The high robustness of both the switch-off character and the conduction windows
is demonstrated and shows the feasibility of the proposed dual structures for real applications. 
\end{abstract}

%61.46.-w, Structure of nanoscale materials
%73.22.-f, Electronic structure of nanoscale materials and related systems
%73.63.-b, Electronic transport in nanoscale materials and structures
%73.23.-b, electronic transport in mesoscopic systems
%72.80.Vp, Electronic transport in graphene
%73.22.Pr, Electronic structure of graphene
%73.23.-b, Electronic transport in mesoscopic systems

\maketitle

\section{INTRODUCTION}
\begin{figure}[t]
  \begin{center}
    \includegraphics[width=8cm]{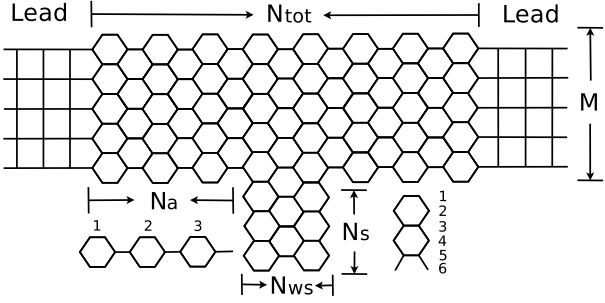}
  \end{center}
  \caption{Schematic view of the armchair GNR with zigzag edged sidearm and
    metal leads. The insets show how the width and length of the specific
    structures are defined. Throughout this work, the width of armchair GNR is
    taken to be $M=21$. } 
  \label{figtw1}
\end{figure}

Ever since the monolayer graphene was first successfully produced experimentally,\cite{first}
intriguing properties from its strictly two-dimensional structure and
massless Dirac-like behavior of low-energy 
excitation have been intensively investigated.\cite{review1,review5}
 Of particular interest are the  
graphene nanoribbons (GNRs) that are strips of graphene obtained by different
methods, e.g., high-resolution lithography,\cite{Kim1,Kim2} chemical
means,\cite{Dai1,Cai1} or most recently the 
unzipping of carbon nanotubes.\cite{Dai2,Kosynkin} Their semiconducting character with a tunable band
gap sensitive to the structural size and geometry makes them good candidates for
future electric and spintronic devices.\cite{Frank} GNRs are classified into two basic
groups, namely, armchair and zigzag ones, according to edge termination
types.\cite{Fujita1,Fujita2,Fujita3}
In the framework of the nearest neighbor tight binding model, the zigzag
  GNRs are always metallic and exhibit special spin-polarized edge
  states.\cite{Fujita1,Fujita2,Fujita3} For the armchair GNRs with 
  width $M$ (as defined in Fig.~\ref{figtw1}), they are metallic when  $M=3n+2$,
  with $n$ being an integer, and semiconducting otherwise.\cite{Fujita2,Fujita3}
Graphene field-effect transistors have been experimentally realized 
by making use of the band gap introduced in GNRs.\cite{Dai1,Dai3} However,
large switching voltage up to several volts is
needed due to the thick back-gate oxides used in these devices. Moreover, the
excellent feature of a tunable band gap in GNRs has not been used in the
overall back-gate configuration.\cite{Dai1,Dai3} 
The electronic transport in GNR-based nanodevices has also be investigated 
theoretically.\cite{Areshkin,Nguyen,Orellana,Jauho}
 Particular energy dependences of
conductance, resulting from the interference effects, are reported in metallic
GNRs\cite{Areshkin,Nguyen} or in semiconducting 
ones out of the gap regime.\cite{Nguyen,Jauho} However, the robustness of these
transport properties against disorder has been shown to be
questionable.\cite{Areshkin,Jauho,Roche,Schubert,Mucciolo}

In this work, we introduce two classes of structures based on gapped armchair
GNRs by using either sidearm or on-site gate voltage, which allow ``on'' and 
``off'' operations in the gap regime. A schematic view of the armchair GNR with 
one sidearm is shown in Fig.~\ref{figtw1}. Other configurations,
i.e., with two sidearms, one or two on-site gate voltages, are shown together
with the numerical results of transport behaviors in the following
figures. Such structures are within the reach of nowaday technology,
  i.e., the patterned GNR can be obtained through high-resolution lithography\cite{Kim1,Kim2} and
the contact\cite{Kim1,Kim2} and top-gate\cite{Liao} technologies are also well developed. It is
noted that throughout this work, the width of the GNR is
taken as $M=21$, so there is a band gap in the pristine
GNR.\cite{Fujita2,Fujita3} Here the corresponding band gap is about
$400$~meV. The two terminals of the GNR are connected to semi-infinite metal
leads,\cite{Nguyen,Martin,Robinson} which simulates the real experimental
condition.\cite{Huard} Such a 
configuration is crucial to access possible states
in the gap regime of the GNR, due to the propagation modes in the metal leads
which are otherwise absent in graphitic ones.\cite{Robinson,Martin}
Meanwhile, the effective length of the sidearm can be electrically
adjusted by a gate voltage (not shown in the figure).\cite{Shen} 
We show that by increasing the length of the sidearm $N_s$, or by increasing the
strength of positive or negative on-site gate voltage in the configuration 
shown in Fig.~\ref{figtw4}(a), conduction peaks are 
introduced into the originally switched-off gap regime.
 The positions of these conduction peaks are determined by the
gate voltage, which, in addition to the common property of switching on and off,
allows us to selectively choose electrons of a particular energy while filter out
the others.  We further propose two
schemes of structures with markedly improved robustness against disorder by
employing two sidearms or on-site gates [see 
Fig.~\ref{figtw5}(a) and (d)]. Due to the
resonance between the two conduction peaks induced by the dual structure, a
conduction plateau, i.e., a broad energy window in which the transmission is close
to one, is formed. This conduction window is very robust against disorder, which
makes our proposal highly feasible for real applications.

\section{Model and Hamiltonian}
We describe the structures consisting of an armchair GNR coupled with metal leads
by using the tight-binding Hamiltonian with the nearest-neighbor approximation, 
\begin{equation}
H=H_L+H_C+H_R+H_T,
\end{equation}
where $H_{L,R}$ are the Hamiltonians of the left and right leads, respectively,
$H_C$ is the Hamiltonian of the GNR and $H_T$ stands for the coupling between
the GNR and the leads.  These terms are written as
\begin{eqnarray}
H_\alpha&=&-t_\alpha\sum_{\langle i_\alpha,j_\alpha
  \rangle}c^\dagger_{i_\alpha}c_{j_\alpha}, \hspace{1cm} \alpha=L,R\\
H_C&=&\sum_{i_c}\varepsilon_{i_c} c^\dagger_{i_c}c_{i_c}-t\sum_{\langle i_c,j_c
  \rangle}c^\dagger_{i_c}c_{j_c},  \label{Eq:GNR}\\
H_T&=&-t_T\sum_{\alpha=L,R}\sum_{\langle i_\alpha,j_c
  \rangle}(c^\dagger_{i_\alpha}c_{j_c}+H.c.) .
\end{eqnarray}
Here, the index $i_c$ ($i_\alpha$) is the site coordinate in the GNR (metal
leads) and $\langle i, j\rangle$ denotes pair of nearest neighbors. $t_\alpha$
and $t_T$ are hopping parameters in the metal leads and between the leads and
the GNR, respectively, which are taken to be equal to the hopping element $t$ in the
GNR.\cite{Nguyen,note1}  The on-site energy in the GNR $\varepsilon_{i_c}$ is
modulated by the on-site gate voltage, which equals $U_g$ in the gated region
and zero elsewhere. 

Within the Landauer-B\"uttiker framework,\cite{Buttiker} the transmission amplitude is given by 
\begin{equation}
T(E)={\rm tr}[\Gamma_L(E)G_{C}^r(E)\Gamma_R(E)G_{C}^a(E)]
\end{equation}
in which $\Gamma_{L/R}$ denotes the self-energy of the isolated ideal leads and
$G_{C}^{r/a}(E)$ represents the retarded/advanced Green's
function for the GNR.\cite{Datta} Here, $E$ is the Fermi energy in the leads.

\section{Results and Discussion}
\subsection{Electronic transport in GNRs with one sidearm or on-site gate voltage}
\begin{figure}[b]
      \includegraphics[width=7cm]{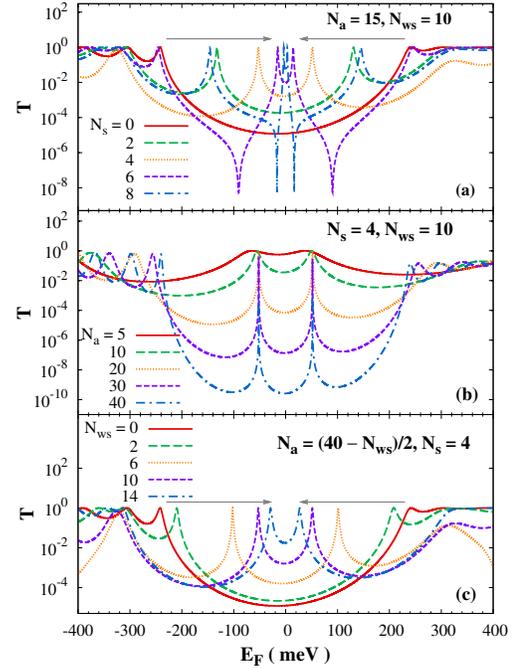}
  \caption{(Color online) Transmission $T$ as function of the Fermi energy of
    in the GNR with one sidearm shown in
    Fig.~\ref{figtw1}: (a) dependence on the length of the sidearm $N_s$. The two gray 
    arrows here and hereafter are plotted to indicate the evolution of
    quantities with the varying of parameters, accordingly; (b) dependence on the total length
    of the GNR $N_a$; (c) dependence on the width of the
    sidearm $N_{\rm ws}$.  All necessary parameters
    are indicated in the corresponding figures.} 
  \label{figtw2}
\end{figure}

\begin{figure}[t]
%    \hspace{-0.65cm}
%  \begin{minipage}[]{5cm}
%      \includegraphics[width=4.5cm]{figtw2a.eps}
%  \end{minipage}
%\vskip -3cm
%    \hspace{-1 cm}
%  \begin{minipage}[]{5cm}
      \includegraphics[width=7cm]{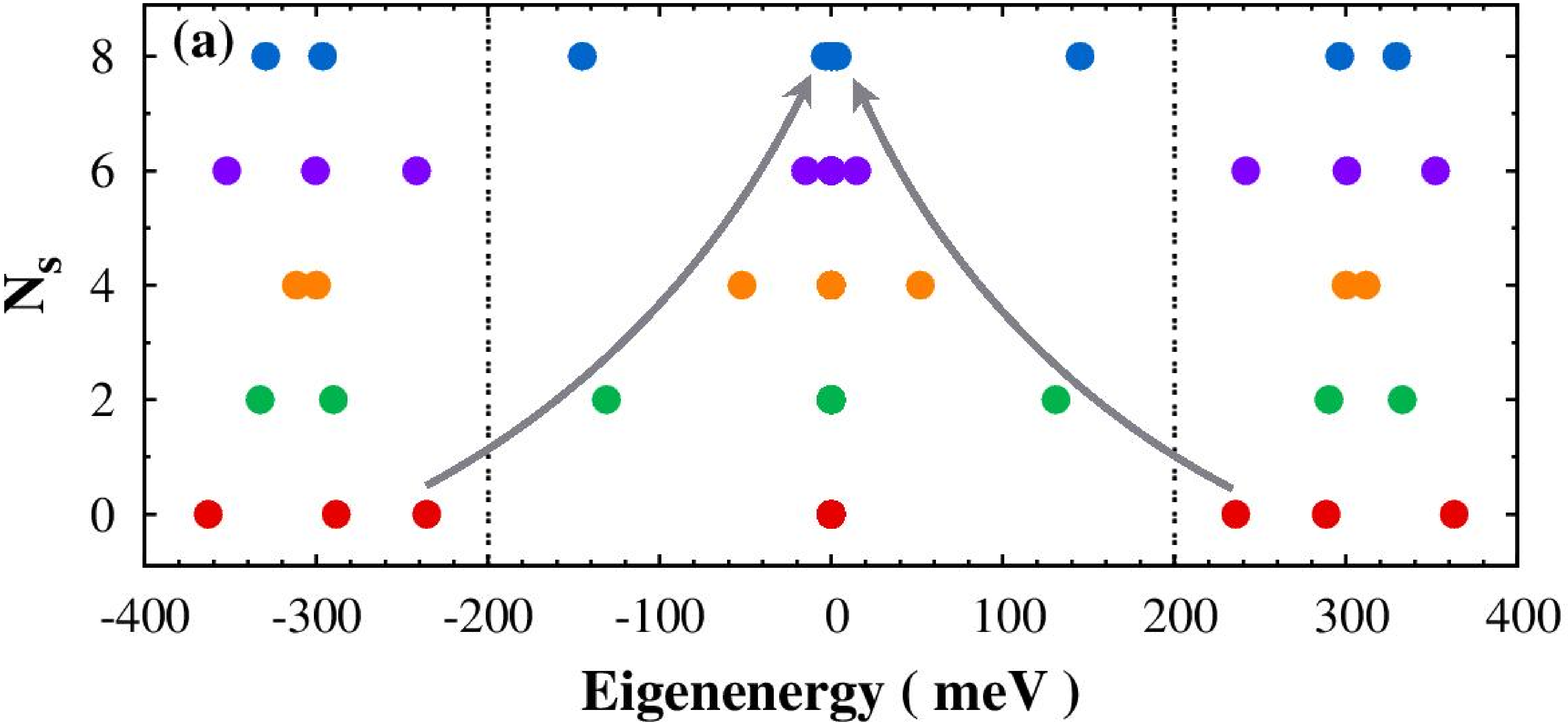}
      \includegraphics[width=8.5cm]{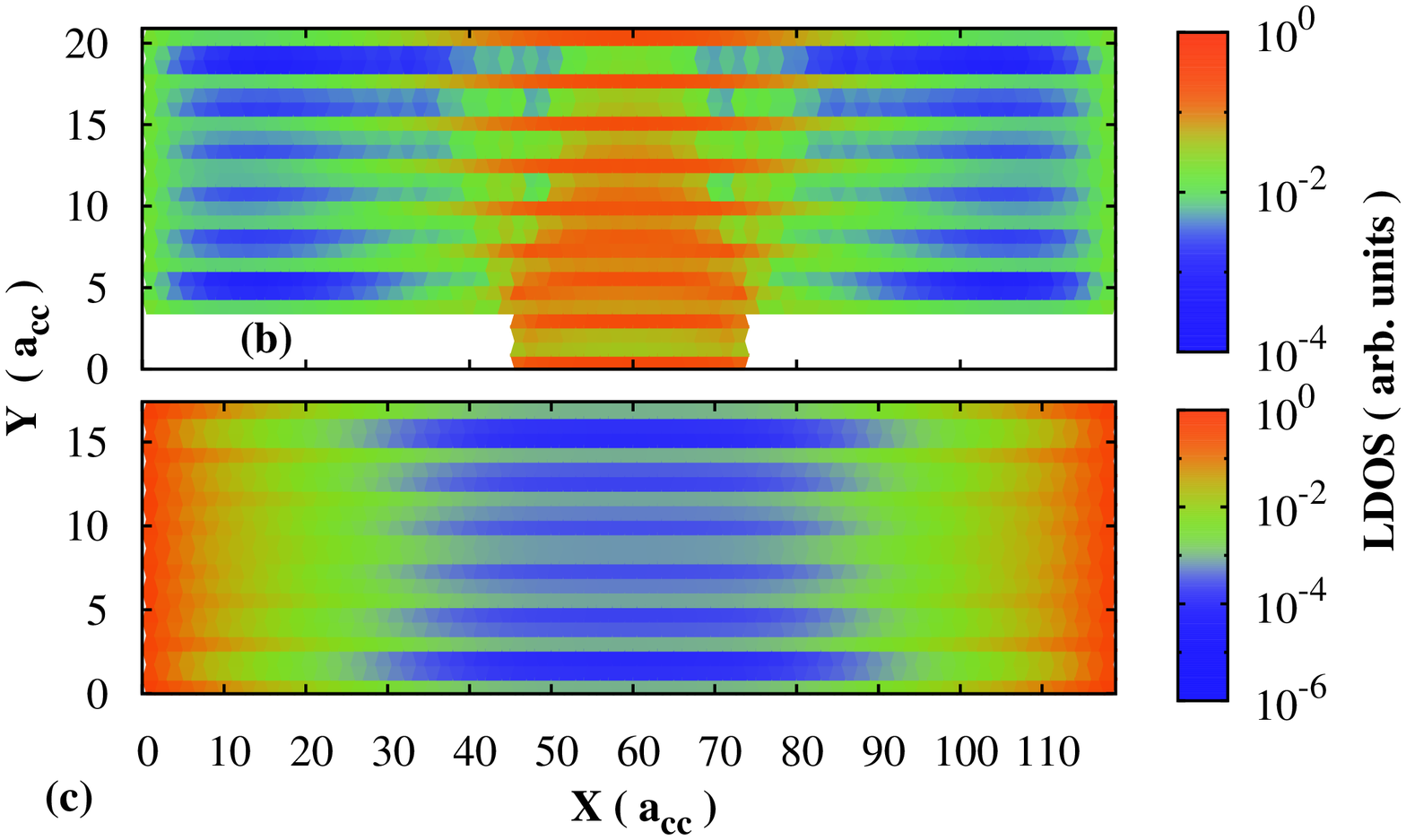}
%  \begin{minipage}[]{10cm}
%    \hspace{-1.5 cm}\parbox[t]{5cm}{
%      \includegraphics[width=4.5cm]{figtw3b.eps}}
%    \hspace{-0.7 cm}\parbox[t]{5cm}{
%      \includegraphics[width=4.5cm]{figtw3c.eps}}
%  \end{minipage}
%  \end{minipage}
  \caption{(Color online) (a) Eigenenergies in the isolated GNR with
    configurations being the same as those in Fig.~\ref{figtw2}(a). The gap regime is indicated
    between the two dashed lines.  Local density of states in the GNR at Fermi energy
    $E_F=51.511$~meV: (b) corresponding to the
    right peak in the gap shown in Fig.~\ref{figtw2}(b) for $N_a=15$, $N_{\rm
      ws}=10$ and $N_s=4$; and (c) in the same condition but without a
    sidearm. $a_{cc}$ is the carbon-carbon bond distance. Note that
    the scale in the $y$-axis is elongated to make the figures clearer and
    the local density of states in the leads are not included.} 
  \label{figtw3}
\end{figure}

We first investigate the transport properties in the GNR with one sidearm as shown
in Fig.~\ref{figtw1}.  In Fig.~\ref{figtw2}(a)-(c), the transmissions are plotted
as function of the Fermi energy. The transmission in the pristine GNR is
indicated by the red solid curve in 
Fig.~\ref{figtw2}(a). It is seen that the band gap manifests
itself in the electronic transport 
behavior that the transmission is well below $10^{-3}$ in the gap regime, i.e.,
$E_F \in (-200, 200)$~meV. This may serve as the ``off'' state of the device
with excellent switch-off character. 
By increasing the length of the sidearm $N_s$ with 
fixed width $N_{\rm ws}=10$, one notices that two conduction peaks from the
positive and negative energy sides are introduced into the gap regime and moving
towards the Dirac point symmetrically. They correspond to n- 
and p-type channels, respectively.
 In order to elucidate this behavior, we calculate the
eigenstates and eigenenergies of the isolated GNRs with configurations employed
in Fig.~\ref{figtw2}(a) and the results are plotted in Fig.~\ref{figtw3}(a). The
eigenenergies are indicated by points with the same 
color as in Fig.~\ref{figtw2}(a) for the corresponding length of the sidearm $N_s$. It
is noted that the states with eigenenergies at the 
Dirac point are the localized edge states in the zigzag
terminals,\cite{Fujita1,Fujita2,Fujita3} which do not really exist when the
terminals are connected to the metal leads. Apart from these fake states, as
indicated by the gray arrows, two states (even more for $N_s \ge 8$) come into the gap and
move towards the Dirac point symmetrically. By comparing Fig.~\ref{figtw2}(a) and
Fig.~\ref{figtw3}(a), close correspondence between the positions of the conduction peaks and
those of the eigenenergies is seen. We hence conclude that the transport behavior in this
structure of a small size can be understood as resonant tunneling through the GNR
via the confined states therein. We then examine how the conduction peaks induced by
the sidearm are influenced by the total length of the GNR. 
From Fig.~\ref{figtw2}(b), one observes that the positions of 
conduction peaks are insensitive to the total length of the GNR.
 This suggests that the states which contribute
to the conduction peaks distribute mainly in the sidearm region. 
Moreover, with the increase of the  total length of the GNR, the confined states
are less coupled to the leads. As a result, the
conduction peaks are narrowed and the 
transmissions in the gap regime other than the conduction peaks are suppressed. 
In addition, as shown in Fig.~\ref{figtw2}(c), a wider
sidearm is more effective in bringing conduction peaks into the gap regime.

It is illustrative to perform a spatial analysis of the conductance. In
Fig.~\ref{figtw3}(b), we plot the local density of states in the GNR 
corresponding to the right conduction peak shown in Fig.~\ref{figtw2}(b) for
$E_F=51.511$~meV, $N_a=15$, $N_{\rm ws}=10$ and $N_s=4$; and in
Fig.~\ref{figtw3}(c) for the same condition but without the sidearm. One
observes that distinct from the case without the sidearm where the 
electronic state is restricted in the vicinities of the two terminals, the
state contributing to the conduction peak indeed mostly distributes in  
the sidearm region.
 This kind of bound states have been discussed by
Sevin\c{c}li {\em et al}.\cite{Sevincli} and Prezzi {\em et al}.\cite{Prezzi}
in the superlattice structures of GNR. 
The underlying physics is that, since the energy
band gap of armchair GNR shows strong dependence on the ribbon width, one can
fabricate structures similar to the conventional semiconductor 
heterojunctions
by joining GNRs with different widths. In the condition shown in
Fig.~\ref{figtw3}(b), segments of armchair GNRs with widths $M=21$, 25 and
21 are joined together. The band gaps of the corresponding infinite GNRs
with the same widths are $(-200, 200)$, $(-140, 140)$ and
$(-200, 200)$~meV, respectively. Therefore, 
the GNR with one sidearm resembles the quantum
well structure in semiconductors with  conduction- and valence-band
offset $\Delta=60$~meV, and hence the bound states are formed therein. 
It is further  noted that due to the finite lengths of the 
GNR segments and the detailed joining
condition, the actual band offset is different from the above simple 
estimation. 

\begin{figure}[t]
      \includegraphics[width=3.2cm]{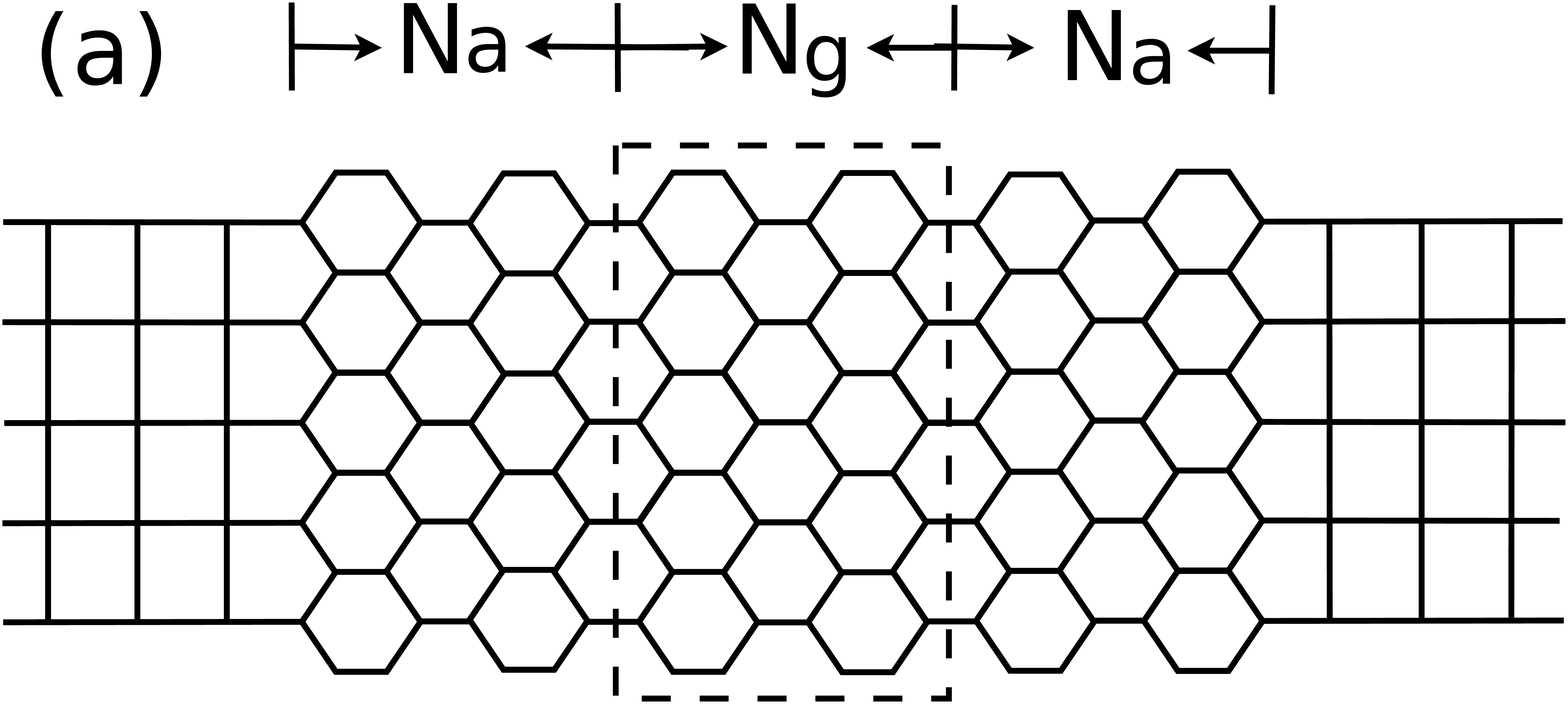}
      \includegraphics[width=5.3cm]{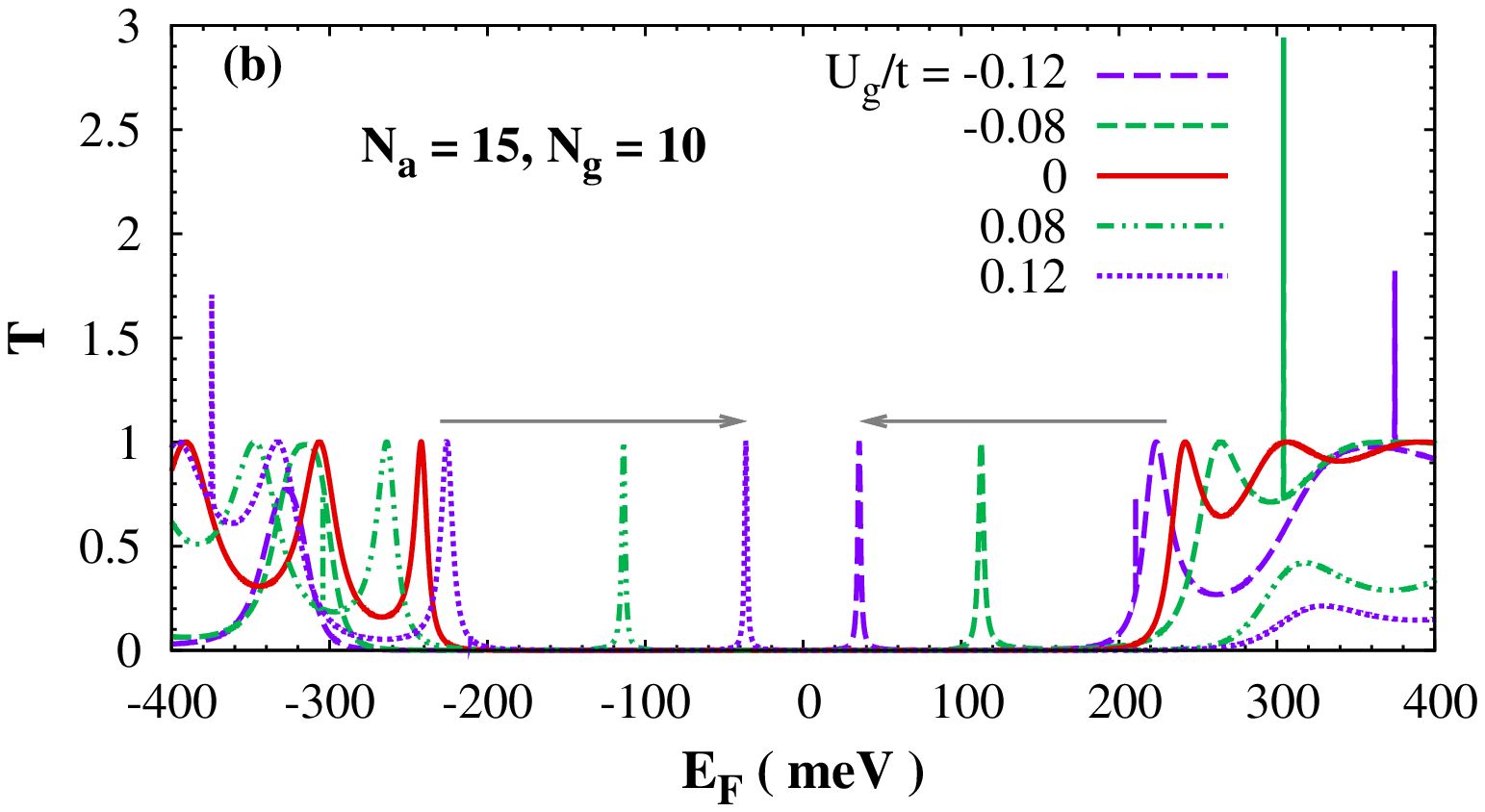}
      \includegraphics[width=8.5cm]{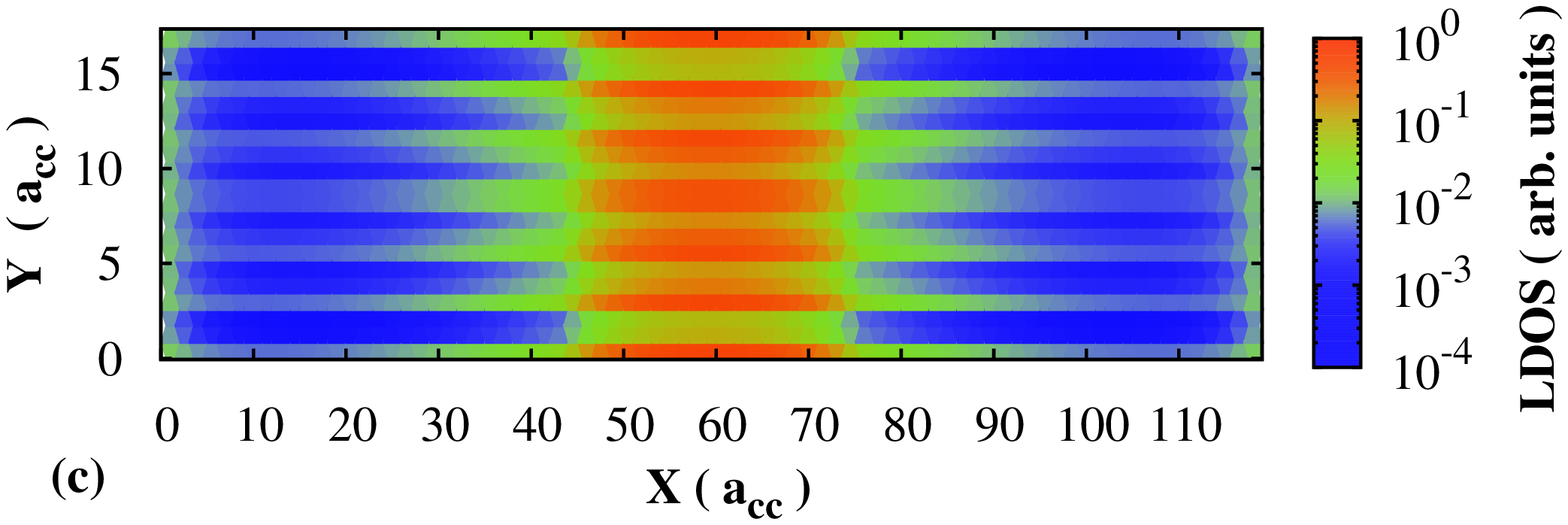}
  \caption{(Color online) (a) Schematic 
    view of the armchair GNR with on-site gate voltage deposited in the region 
    labelled with dashed box. (b) Transmission $T$ as function of the Fermi
    energy of in the GNR with a gate voltage shown in
    (a), for different values of on-site energy $U_g$. (c) Local density of
    states in the GNR corresponding to the
    conduction peak induced by an on-site gate voltage $U_g=-0.12t$ in (b), at
    Fermi energy $E_F=33.405$~meV.  All necessary parameters
    are indicated in the corresponding figures.} 
  \label{figtw4}
\end{figure}

Following the idea of introducing bound states into the gap regime, we then
propose another way of accessing states inside the band gap by using an  on-site gate
voltage.\cite{Silvestrov,Recher} As illustrated in Fig.~\ref{figtw4}(a), we apply a
positive (negative) voltage in the framed region by a top gate\cite{Liao}
which acts as a well potential to electrons (holes). The transmissions as function
of the Fermi energy are plotted in Fig.~\ref{figtw4}(b) for different values of 
the gate voltage. One observes that by increasing the strength of positive (negative) gate
voltage, a conduction peak enters the gap regime from the right (left) and moves
towards the left (right). 
These conduction peaks are from the tunneling via the 
bound states in the gapped region.\cite{Silvestrov} We demonstrate this by
plotting the local density of states in
Fig.~\ref{figtw4}(c), which corresponds to the conduction peak induced by a gate
voltage $U_g=-0.12t$.
 In the sense that here one can introduce only one
conduction peak into gap regime with its position fully determined by the gate voltage, this configuration
has the advantage to serve as an energy filter.  The influences of the width of
the gate region  and the total length of the GNR on the induced conduction peaks
resemble the case with one sidearm, and are not explicitly plotted.

\begin{widetext}
\begin{figure}[b]
  \begin{minipage}[]{20cm}
    \hspace{-2.5 cm}\parbox[t]{6cm}{
      \includegraphics[width=4.8cm]{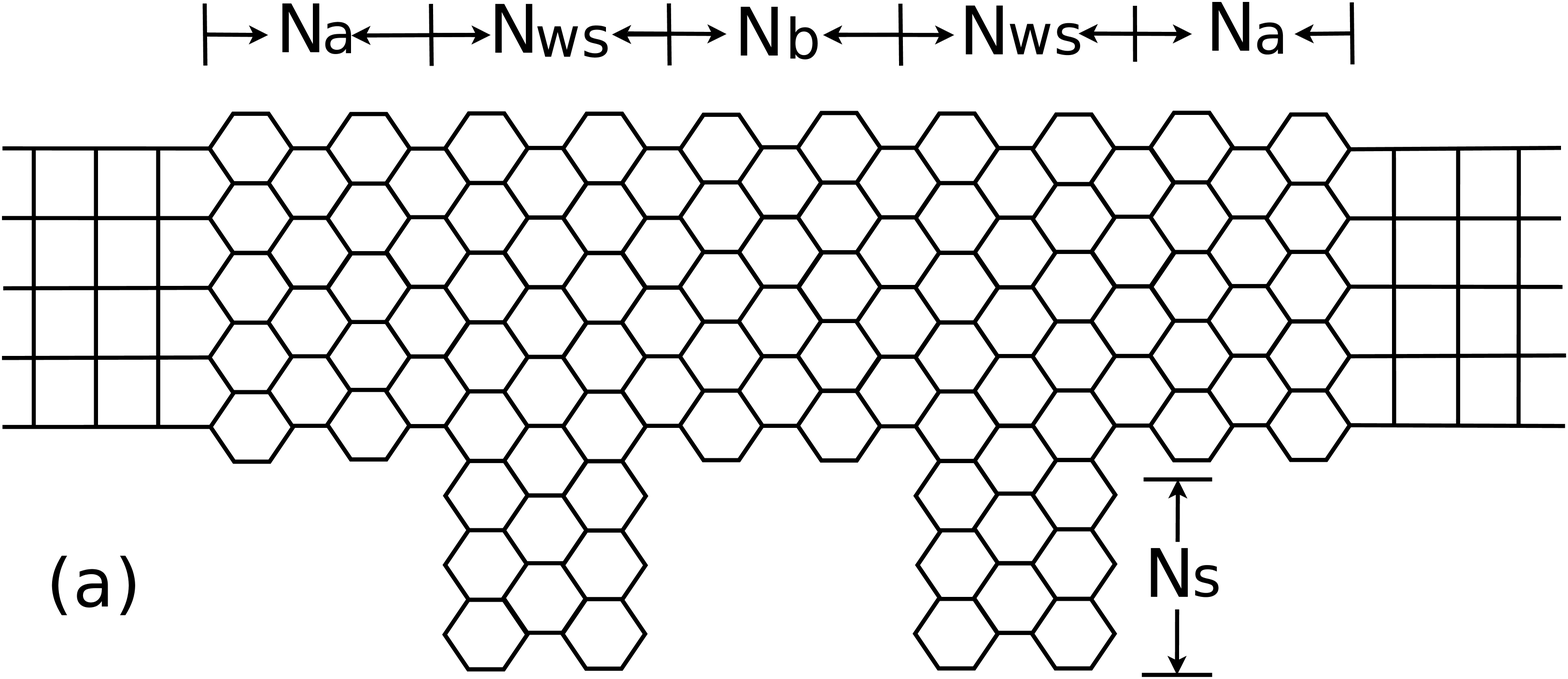}}
    \hspace{-0.5 cm}\parbox[t]{6cm}{
      \includegraphics[width=5.8cm]{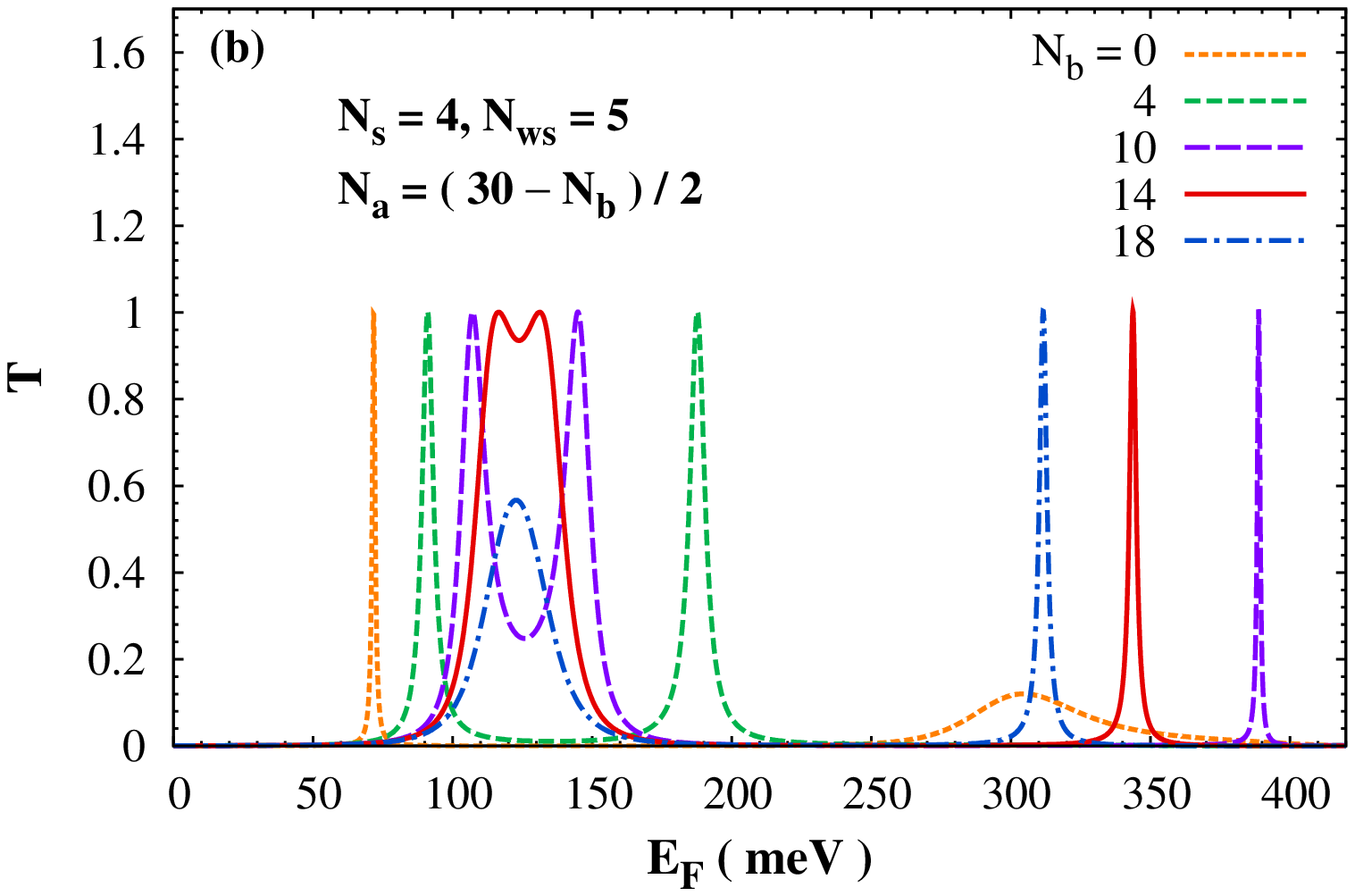}}
    \hspace{-0.5 cm}\parbox[t]{6cm}{
      \includegraphics[width=5.8cm]{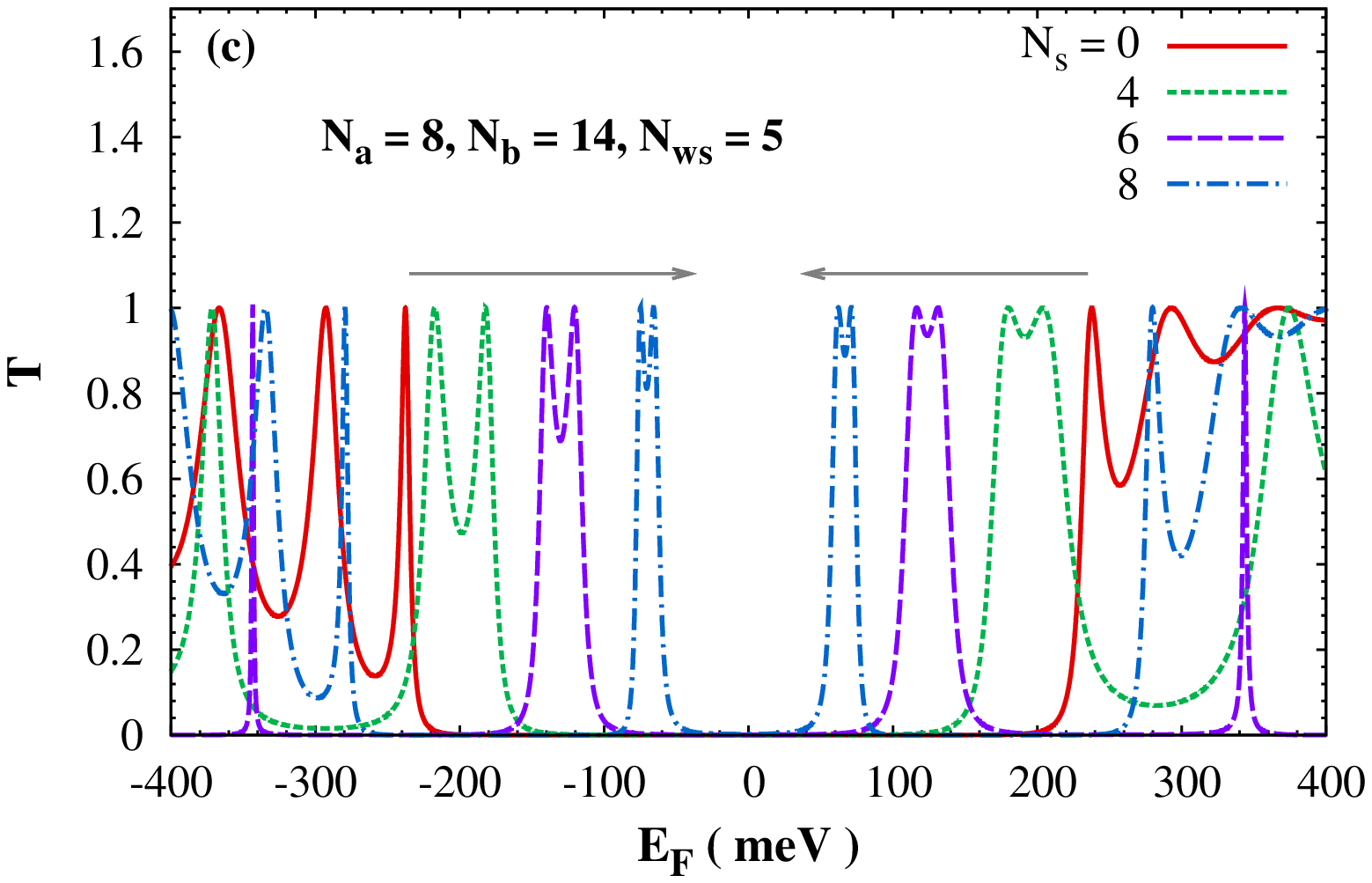}}
  \end{minipage}
  \begin{minipage}[]{20cm}
    \hspace{-2.5 cm}\parbox[t]{6cm}{
      \includegraphics[width=4.8cm]{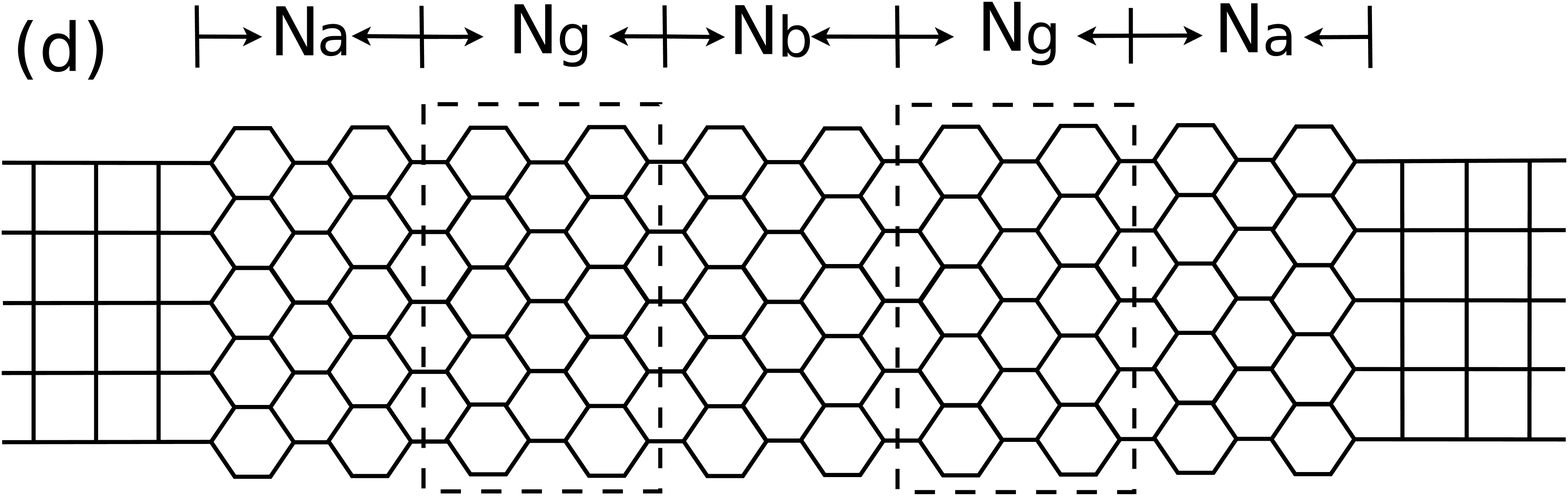}}
    \hspace{-0.5 cm}\parbox[t]{6cm}{
      \includegraphics[width=5.8cm]{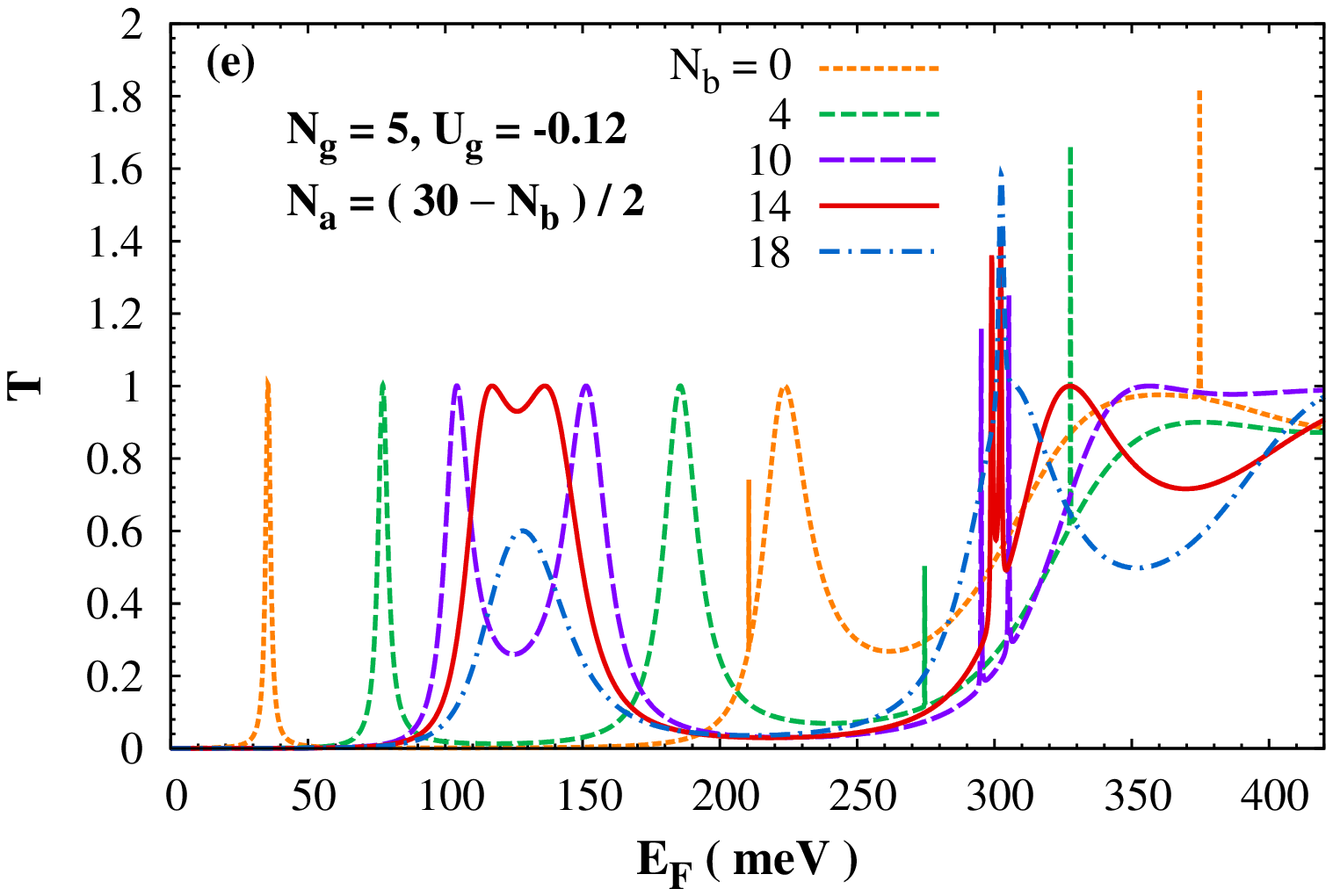}}
    \hspace{-0.5 cm}\parbox[t]{6cm}{
      \includegraphics[width=5.8cm]{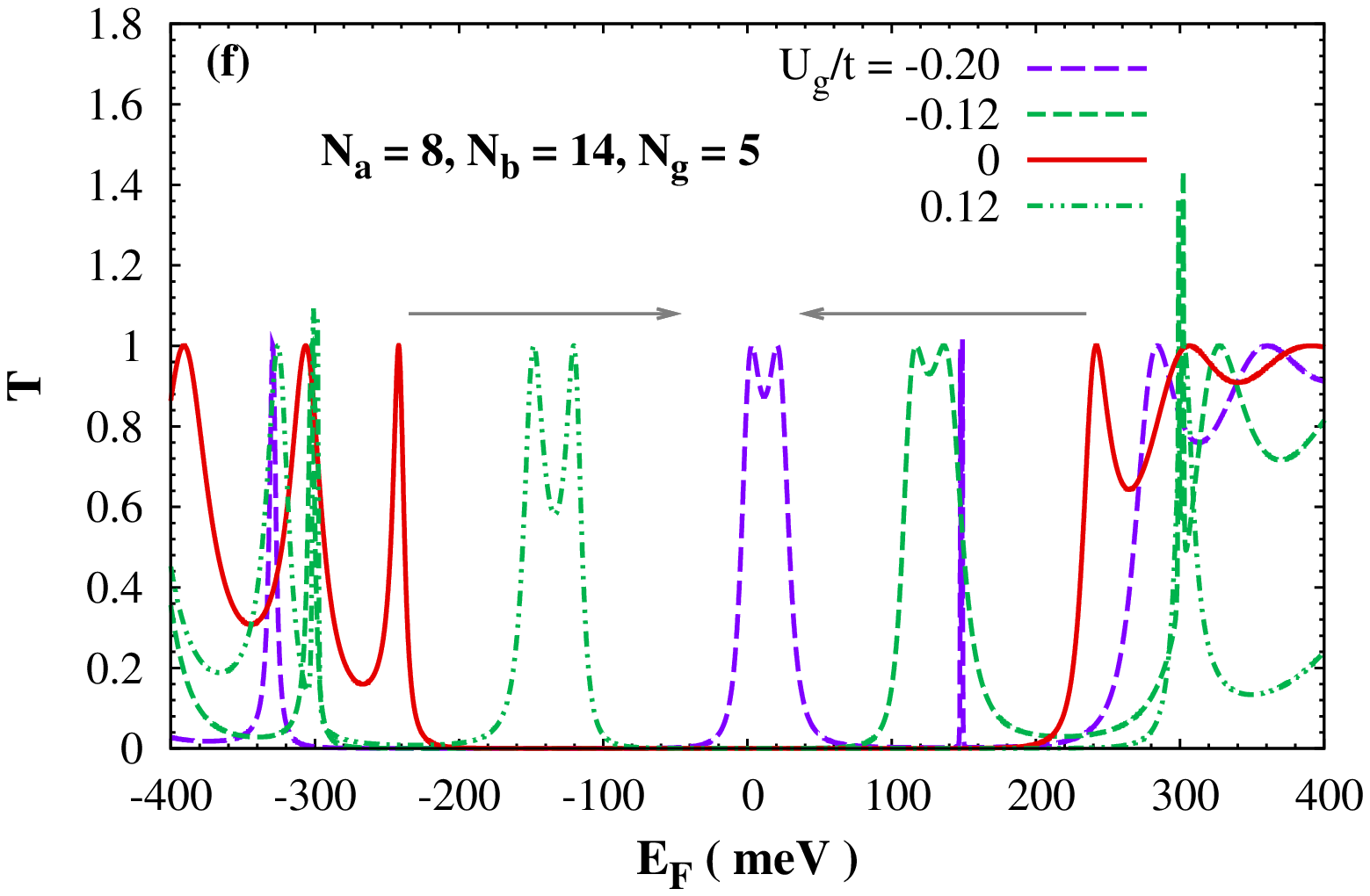}}
  \end{minipage}
  \begin{minipage}[]{17.5cm}
  \begin{center}
  \caption{(Color online) (a) and (d) Schematic
    view of the armchair GNR with two sidearms and two on-site gates,
    respectively. Transmission $T$ as function of the Fermi energy 
    in the GNR shown in (a): (b) dependence on the spacing 
    between the two sidearms $N_b$; (c) dependence on the length
    of the sidearms $N_s$. Transmission $T$ as function of the Fermi energy 
    in the GNR shown in (d): (e) dependence on the spacing 
    between the two gates $N_b$; (f) dependence on the values of on-site energy $U_g$ induced
    by the gate voltage. All necessary parameters
    are indicated in the corresponding figures.} 
  \label{figtw5}
  \end{center}
  \end{minipage}
\end{figure}
\end{widetext}
\subsection{Electronic transport in GNRs with two sidearms or on-site gate voltages}

Due to the fact that the conduction peaks introduced into the gap regime are
extremely sharp  in the above two configurations and hence are easily destroyed by disorders
as we will show below, we further propose two schemes of structures to improve 
the robustness, i.e., by employing two sidearms or gate voltages shown in
Fig.~\ref{figtw5}(a) and (d).\cite{Shen,Wang,Bliokh,Michael} It is noted that here we fix the total length of
the GNRs $N_{\rm tot}=40$ and the total width of the two sidearms or the gates
is set to be the same width of the sidearm or gate region in the previous configurations.
In searching for the best performance of the devices, i.e., a  
wide conduction window, we vary the
spacing length $N_b$ between the two sidearms  or gate 
regions to modify the interference between tunnelings via the two (quasi-)bound
states from the dual structure. The results are plotted in Fig.~\ref{figtw5}(b)
and (e), for the situations with two sidearms and two gates, respectively. It is
found that for both cases with $N_b=14$, a conduction plateau is formed with a
wide energy window up to 50~meV 
centered around $E_F=125$~meV (red solid curves in the figure). Such a wide conduction
window in the original gate 
region allows a large current in the ``on'' state of the proposed device
which is of potential use for high performance field-effect
transistors. Moreover, as shown in Fig.~\ref{figtw5}(c) and (f), the positions
of the conduction plateaus can be controlled by the length of the sidearms and by
on-site gate voltage, respectively. So the excellent feature of controlled
modification of the conduction window is preserved in the dual structures.

\subsection{Disorder analysis}

\begin{figure}[t]
%  \begin{minipage}[]{10cm}
      \includegraphics[width=7cm]{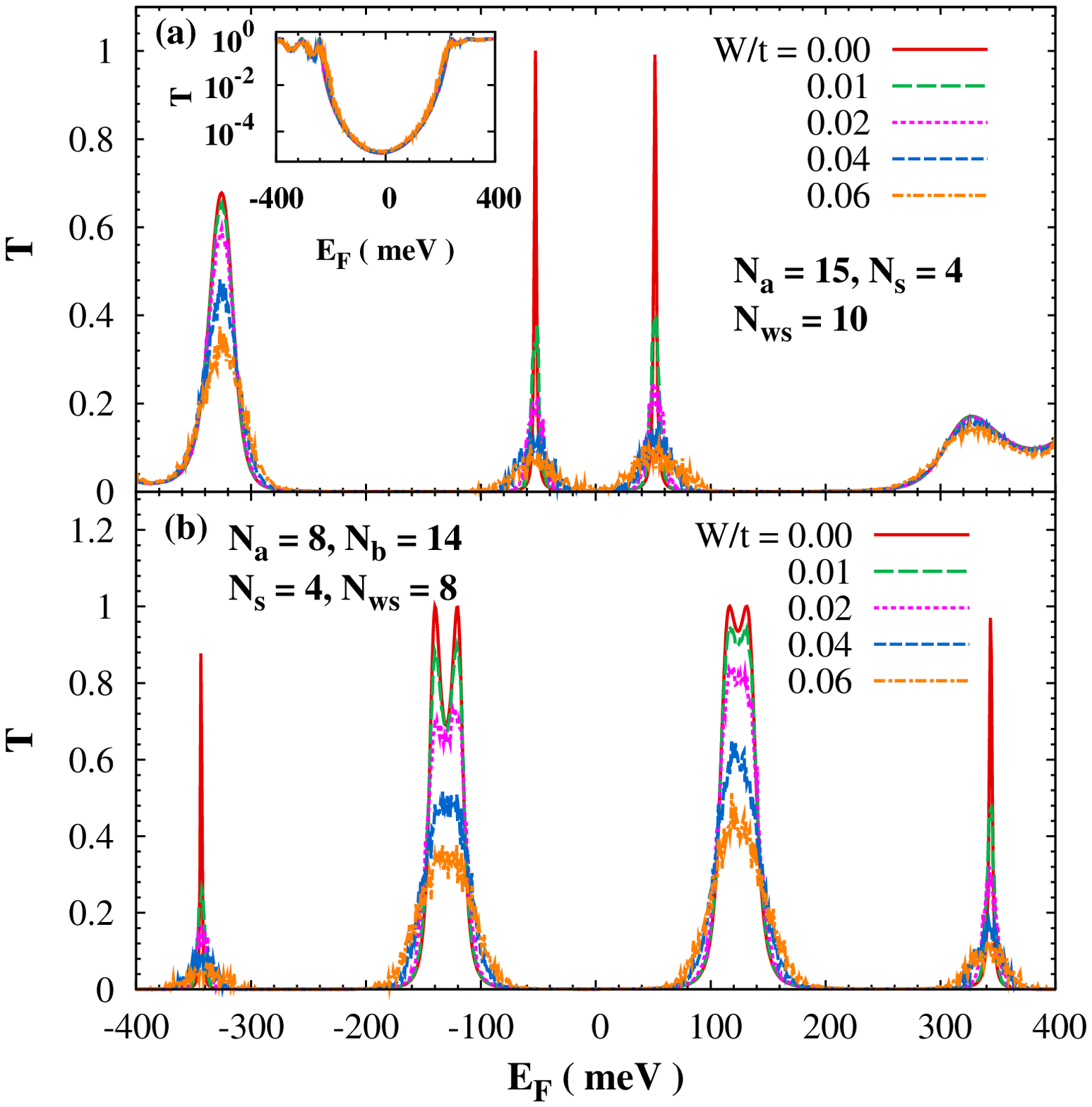}\\
      \includegraphics[width=7cm]{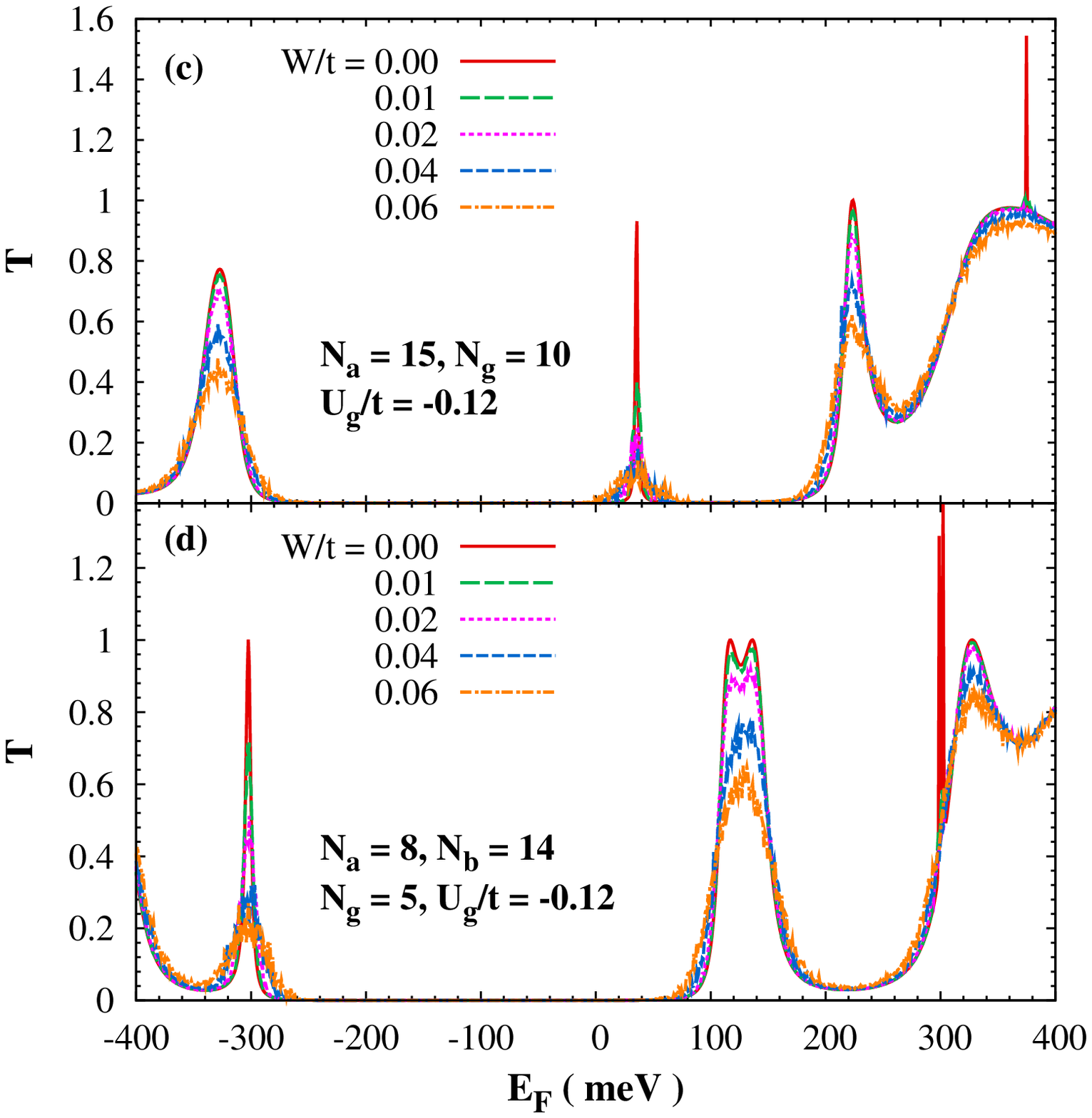}
%  \end{minipage}
  \caption{(Color online) Transmission $T$ as function of the Fermi
    energy with different Anderson disorder strength $W$ in the GNR: (a) and (b) with
    one and two sidearms [corresponding to GNR structures shown in
    Fig.~\ref{figtw1} and Fig.~\ref{figtw5}(a)], respectively; (c) and (d) with
    one and two one-site gate voltages [corresponding to GNR structures shown in
    Fig.~\ref{figtw4}(a) and Fig.~\ref{figtw5}(d)], respectively. All necessary parameters
    are indicated in the corresponding figures.}  
  \label{figtw6}
\end{figure}
We now show the feasibility of the above proposed devices for real application
by analyzing the robustness of the switch-off character, and more importantly the
conduction peaks and plateaus against the Anderson disorder.\cite{Schubert,Shen} In our
simulation, the Anderson disorder is created out by introducing random on-site
energy at the carbon atoms: 
$\varepsilon^\prime_{i_c}=\varepsilon_{i_c}+\lambda W$ [see Eq.~(\ref{Eq:GNR})]. Here $W$ is the
disorder strength and $\lambda$ is a random number with a uniform probability
distribution in the range $(-1, 1)$. The converged
transmissions are obtained by averaging over 100 random
configurations.\cite{Shen} The
results of pristine GNR without a sidearm or on-site gate voltage are plotted in the inset
of Fig.~\ref{figtw6}(a). One notices that under different strength of disorder,
all five curves almost coincide with each other. This indicates that the gap behavior is very
robust against disorder, which ensures extremely small leakage current in the
``off'' state of the device. We then turn to check the robustness of the ``on''
state of the proposed structures. 
By comparing the corresponding curves in
Fig.~\ref{figtw6}(a) and (b), one finds that the conduction peaks from one
sidearm rapidly decrease with the strength of the disorder whereas the conduction
plateaus from two sidearms are more sustained. The latters are only reduced by about
$50\%$ for the largest disorder strength $W=0.06t$.
 Therefore, the robustness is
immensely improved by using two sidearms to introduce a wide conduction window into the
gap regime. The situation for
configurations with one or two gate voltages is similar. So we only 
plot our results in
Fig.~\ref{figtw6}(c) and (d) without more discussions. In this way, we
demonstrate the robustness of the proposed devices for both ``on'' and ``off''
states.

\section{Summary}
In summary, we have proposed two schemes for field-effect transistor, which may
also work as energy filter, by studying transport properties in the GNR-based
structures with sidearms or on-site gate voltages. Gapped armchair GNRs 
are employed with the band gap used as a natural ``off''
state of the transistor. Metal leads are employed so that by further
introducing a sidearm or on-site gate voltage to the GNR, one is able to 
access the gap regime with conduction peaks. Moreover, by employing two sidearms
or on-site gate voltages, we
obtain much wider conduction windows with the transmission close to one, which
allows a large ``on'' current.  We show that the positions of the conduction
peaks or plateaus can be controlled by the length of the sidearm (which can be modulated by a
gate voltage), or by the voltage of the on-site gates on the GNR. This property
enables the proposed devices not only serve as a common transistor with large on/off ratio, but
also as an energy filter.
We further demonstrate the  robustness of both the ``off'' and ``on'' states of the
devices against disorder. The excellent switch-off ability, the 
wide conduction window of ``on'' state which allows controlled modifications 
and the high robustness against disorder suggest that the proposed structures have great
potential to work as high performance 
field-effect transistor in reality.

\begin{acknowledgments}
This work was supported by the National Basic Research 
Program of China under Grant No. 2012CB922002 and 
the National Natural Science Foundation of China under Grant No. 10725417.
\end{acknowledgments}


\begin{thebibliography}{0}
\bibitem{note1}As shown in Ref.~\onlinecite{Nguyen}, a change in the 
  ratio $t_\alpha/t$ or in the value of $t_T$ does not qualitatively affect the
  overall features of transport. Here we take $t_\alpha = t_T = t$,
  corresponding the case with ballistic tunneling in 
  the GNR-lead interface.\cite{Robinson,Martin}
\end{thebibliography}
\end{document}